# Designing Multi-Step Action Models (MSAMs) for Enterprise AI Adoption

Shreyash Mishra, Shrey Shah, Rex Pereira

*Empsing, 128 City Road, London EC1V 2NX.*

hello@empsing.com

**ABSTRACT** –

This paper introduces the Multi-Step Action Model (MSAM), a closed-source AI model designed by Empsing to address challenges hindering AI adoption in enterprises. Through a holistic examination, this paper explores MSAM's foundational principles, design architecture, and future trajectory. It evaluates MSAM's performance via rigorous testing methodologies and envisions its potential impact on advancing AI adoption within organizations. By providing insights into MSAM's capabilities, this paper offers valuable guidance for organizations seeking to harness AI-driven technologies for sustainable growth and competitiveness in the digital era through Empsing's Platform.

**ABOUT THE AUTHORS –**

Shreyash Mishra, as the CEO of Empsing, sets the strategic direction for the company, focusing on pioneering innovations like the Multi-Step Action Model (MSAM) to propel enterprise AI adoption forward.

Shrey Shah, Empsing's Product Lead, drives the development of MSAM, ensuring its alignment with user needs and market demands through intuitive design and technical expertise.

Rex Pereira, Empsing's Global Head of Business and Adoption, spearheads the global adoption of MSAM, forging strategic partnerships and facilitating its seamless integration into diverse enterprise ecosystems for maximum impact.

## I. INTRODUCTION

In the contemporary landscape of enterprise operations, the integration of Artificial Intelligence (AI) technologies has emerged as a pivotal driver of innovation, efficiency, and competitiveness. Leveraging AI-powered solutions, organizations seek to optimize processes, enhance decision-making, and unlock new avenues for growth and value creation. However, despite the considerable advancements in AI research and development, the adoption of AI within enterprise settings remains fraught with challenges and complexities.

This paper delves into the design and implementation of the Multi-Step Action Model (MSAM), a pioneering AI model developed by Empsing to address the inherent barriers to AI adoption in enterprise environments.

Through a comprehensive examination of the challenges inhibiting AI integration, coupled with the development of a robust foundational framework, MSAM offers a transformative approach to intelligent automation and problem-solving within organizations of all sizes.

Drawing upon empirical research, industry insights, and technological innovations, this paper explores the foundational principles underpinning MSAM, elucidates its design architecture, and evaluates its performance through rigorous testing methodologies. Furthermore, it examines the future of MSAM, envisioning its evolution as a catalyst for innovation, collaboration, and responsible AI deployment in the digital era.

## II. PROBLEM MAPPING

Empsing, in collaboration with its parent company – Cosdec Alpha Group, conducted a comprehensive survey among the enterprise clientele of Cosdec Alpha to delineate the prevailing challenges impeding AI Adoption. Drawing upon Empsing's extensive assessment and augmenting it with insights gleaned from a survey orchestrated by Predibase among organizations experimenting with Large Language Models (LLMs), our inquiry engaged over 200 executives, data scientists, machine learning engineers, developers, and product managers, representing diverse enterprises across more than 20 countries [1]. Systematic analysis of these findings has led to the following discernments:

**Low Accuracy:** The deployment of Generalist language models often yields sub-standard accuracy, particularly evident in specialized tasks. Organizations contend with meeting their accuracy benchmarks when employing generic LLMs due to the opacity inherent in identifying model hallucinations and discerning biased outputs [2].

**Cost and Governance Constraints:** Variability in infrastructure preparedness and compute budget allocation across different enterprises presents formidable challenges during the operationalization phase of AI integration. Pre-packaged LLM solutions are encumbered with substantial financial implications, further exacerbated by enterprise use cases necessitating the handling of sensitive or confidential data, thereby rendering direct use of public LLM APIs impracticable [2].

**Utilizing Enterprise Information:** A prevailing notion among organizations posits that the fine-tuning of private large language models with enterprise-specific content enhances performance. This entails the strategic prioritization of enterprise content to alleviate challenges associated with enterprise information RAG. A notable proportion of teams intend to tailor their LLMs through fine-tuning (32.4%) or reinforcement learning bolstered by human feedback (27%). However, impediments persist, encompassing data scarcity (21%) and infrastructural intricacies (46%) [1]. Nonetheless, the adoption of such strategies is fraught with inherent complexities:

<u>Sustaining Enterprise Relevance:</u> Modifying the focus of LLMs towards enterprise-specific content is fraught with challenges owing to the extensive training data underpinning these models. Consequently, responses may exhibit a confluence of enterprise-centric and pre-existing data, thus complicating the task of ensuring contextual relevance.

<u>Hallucinatory Responses:</u> In the absence of enterprise-specific data pertaining to a given query, LLMs may fabricate responses, engendering outcomes characterized by unpredictability albeit often perceived as credible [3].

**Limited Reasoning:** While LLMs exhibit prowess in text generation tasks such as copywriting, their efficacy in predictive tasks like modelling, forecasting or classification trails behind models specifically tailored for such purposes. Endeavours to leverage LLMs for automating tasks necessitating human-like reasoning and planning encounter constraints, as these models, notwithstanding their capacity to generate plausible-looking reasoning, lack genuine reasoning and planning faculties [4].

## III. FOUNDATION FOR A MULTI-STEP ACTION MODEL (MSAM)

MSAM, or the Multi-Step Action Model, is an AI model methodically engineered for task-based actions by Empsing to cater to diverse enterprise use cases. Serving as the backbone of the Empsing Platform — an AI-based Digital Employees Service — MSAM is designed to decide, act and execute diverse tasks within enterprise environments at par or beyond human accuracy. It draws upon a combination of Large Language Models (LLMs), Machine Learning (ML) Models, Custom Software Modules, Connected Organization Databases, Authorized Connected Apps, Empsing's Base Consciousness System (EBCS), and a Dynamic, Responsible Knowledge Application Protocol. Through this amalgamation of technologies and methodologies, MSAM empowers the Empsing platform and it's digital employees to optimize workflows, streamline processes, and drive innovation for organizations while ensuring reliability, scalability, and ethical application of AI capabilities. Empsing's Multi-Step Action Model (MSAM) is underpinned by succinct foundational elements designed to tackle the complexities of enterprise AI adoption.

**Dynamic Fine-Tuning:** MSAM employs a near real-time, dynamic fine-tuning mechanism to adapt to evolving enterprise data. It covers static (eg. Files) and dynamic (eg. databases) data types. This adaptive agility ensures precise customization without introducing information bias, enhancing MSAM's effectiveness across diverse operational contexts.

**Multimodal Support**: MSAM features robust support for multimodal inputs, including images, videos, URLs, and various document formats. Leveraging multiple AI architectures and models, MSAM processes heterogeneous data sources with unparalleled accuracy, enhancing versatility and applicability in enterprise settings.

**Data Privacy and Security**: MSAM integrates multi-layer encryption protocols and access controls to safeguard sensitive information via the Empsing Watchdog at the model level. By encrypting data at rest and in transit, MSAM ensures compliance with regulatory requirements while maintaining data integrity and confidentiality building privacy by design.

**Modular Integration:** MSAM can adopt and integrate as a modular framework via APIs and Packages for seamless interoperability with existing IT infrastructures. Additionally, Standalone infrastructure components enable flexible deployment – As a Service and On-Premise, scalability, and customization, ensuring minimal disruption to business workflows.

**Knowledge Application Focus:** MSAM prioritizes knowledge application over data accumulation, leveraging advanced techniques to extract actionable insights from diverse data sources. This approach empowers organizations to leverage advanced reasoning capabilities of Empsing's Digital Employees and make informed decisions with precision and confidence without exposing Enterprise Data for training models.

## IV. DESIGNING THE MSAM

The design of Empsing's Multi-Step Action Model (MSAM) is a harmonious integration of advanced AI methodologies and architectural considerations, precisely structured to address the intricate challenges inherent in enterprise environments. MSAM's adaptability and flexibility are conceptually represented as a dynamic neural network, facilitating efficient resource allocation, node management and adaptation to varying task complexities and data availabilities.

**Embedded in a Multi-Tenancy Architecture**: MSAM works atop a multi-tenancy architecture, engineered to ensure scalability, security and flexibility across varied deployment landscapes. This architectural foundation enables Empsing to seamlessly adapt MSAM to the diverse infrastructural nuances of enterprises, facilitating isolation and security between tenants while accommodating dynamic scaling requirements.

**Data Processing and Dynamic Fine-Tuning Mechanisms**: Core to MSAM's design is a sophisticated framework for data processing and dynamic AI fine-tuning, engineered to harness organizational data for real-time adaptation and optimization. Leveraging multitude of data & AI algorithms, this mechanism empowers MSAM to refine its pathways continuously, ensuring precision and relevance in task execution across evolving operational contexts. In addition, the Empsing Watchdog system allows to monitor queries and responses from preventing leak of sensitive information, knowledge hacking, poisoning or inappropriate language usage.

**AI Component Integration**: MSAM integrates a myriad of AI components such as Large Language Models (LLMs), Machine Learning (ML) Models, Custom Software Modules, Connected Organization Databases, Authorized Connected Apps, Empsing's Base Consciousness System (EBCS), and a Dynamic, Responsible Knowledge Application Protocol. It also harnesses an array of AI methodologies on-demand, as needed, including Evol-Instruct, Reinforced Evol-Instruct, Mixture of Experts, etc., augmenting its cognitive capabilities and adaptability to dynamic task requirements. This cohesive integration of AI components & techniques fosters interoperability and synergy, enabling MSAM to leverage their collective capabilities effectively.

**Customized Task Resolution Flow:** Diverging from conventional language models, MSAM embodies a task-based action approach specifically designed for planning, executing, and monitoring tasks within enterprise contexts. MSAM features a tailored query resolution flow embedded in a virtual workspace, facilitating dynamic selection of components and systems to address queries and tasks with precision and efficiency. This adaptive approach optimizes task execution by aligning MSAM's outputs with enterprise-specific needs and preferences, enhancing overall performance and user satisfaction.

**Three Phases of Task Execution**: Orchestrating task execution within MSAM are three distinct phases: a. Planning, b. Decision-making, and c. Executing & Monitoring Actions. Each phase is arranged to optimize task execution, ensuring seamless coordination and alignment with organizational objectives while fostering operational agility and responsiveness.

**Ethical AI Practices**: MSAM upholds the highest standards of ethical AI practices, guided by principles of fairness, transparency, and accountability. Rigorous measures, through Empsing Watchdog, such as bias mitigation, explainability, and algorithmic transparency are embedded within MSAM's design, mitigating the risk of unintended consequences and fostering trust in its deployment.

**V. TESTING METHODOLOGY**

Empsing's assessment of the Multi-Step Action Model (MSAM) for enterprise readiness was underpinned by a meticulously structured testing methodology aimed at scrutinizing its efficacy across diverse operational dimensions. This section elucidates the testing procedures employed, emphasizing their rigor and focus on evaluating MSAM's adaptability and performance within enterprise environments.

**User-Centric Evaluation:** The initial phase of assessment entailed a supervised selection of 300+ individuals from internal teams, partner organizations, and testing entities, representing a diverse spectrum of roles and expertise levels within enterprises. This cohort was stratified into three distinct categories: Specialists, Generalists, and Observers, mirroring the varied personas encountered in enterprise settings. Specialists, with their domain-specific proficiency, Generalists, involved in interdisciplinary operations, and Observers, assuming managerial oversight, constituted a comprehensive cross-section reflective of enterprise dynamics.

Each participant was granted access to the Empsing platform for a minimum duration of two months, during which they were tasked with engaging in a typical workflow involving the submission of a minimum of 1,000 queries/tasks and active involvement in at least 10 projects monthly. This exhaustive regimen was designed to simulate & embed real-world enterprise scenarios into testing, facilitating a thorough evaluation of MSAM's performance across a multitude of operational contexts, ranging from specialized tasks to cross-functional projects. Totalling over 400,000 requests and more than 300 projects in 2 months, MSAM was tested by processing over 15 billion tokens of text and other modalities.

**Benchmark Assessment**: Complementing the user-centric evaluation, MSAM underwent rigorous benchmark-driven assessments to gauge its performance against established standards and industry benchmarks indicative of enterprise readiness. The selection of benchmarks was carefully curated to encompass diverse facets of MSAM's functionality, with a particular focus on multi-step reasoning capabilities and proficiency across various domains inherent to enterprise operations.

AGIEval: Positioned as a human-centric benchmark, AGIEval leverages standardized examinations, including college entrance exams and professional qualification tests, to assess MSAM's proficiency in tasks aligned with human cognition and problem-solving. By benchmarking MSAM's performance against average human performance and other models on these exams, AGIEval provided valuable insights into its cognitive capabilities and readiness for enterprise deployment.

HumanEval: Designed as a comprehensive evaluation of MSAM's programming prowess, HumanEval comprised a curated set of programming problems spanning algorithmic complexity, mathematical reasoning, and creative problem-solving. The benchmark scrutinized MSAM's ability to generate code and articulate solutions in natural language, thereby assessing its suitability for addressing diverse programming needs in enterprise contexts.

GAIA: Serving as a litmus test for MSAM's multimodal capabilities to perform as a General AI Agent, GAIA presented a series of real-world questions necessitating multifaceted reasoning, multi-modality handling, and adept problem-solving skills. GAIA is made of more than 450 non-trivial question with an unambiguous answer, requiring different levels of tooling and autonomy to solve. It is therefore divided in 3 levels, where level 1 should be breakable by very good LLMs, and level 3 indicate a strong jump in model capabilities. By challenging MSAM with tasks analogous to those encountered by human counterparts, GAIA provided critical insights into its readiness to navigate complex enterprise scenarios and human-level performance.

By simulating real-world workflows and benchmarking against established standards, Empsing ensured a comprehensive evaluation of MSAM's enterprise readiness, laying a robust foundation for its deployment in diverse organizational settings.

**VI. USER-CENTRIC TESTING RESULTS**

Empsing's User-Centric Evaluation, conducted over a two-month period, encompassed a comprehensive analysis of over 400,000 requests and more than 300 projects, processing a vast corpus of text and multimodal data exceeding 15 billion tokens. The findings reveal strong performance metrics (human-reported) across all evaluated parameters:

| Metric | CS | CG | CO | Avg. |
|---|---|---|---|---|
| Task Accuracy | 87% | 94% | 91.25% | 90.75% |
| Task Completion | 91% | 96.5% | 90% | 92.5% |
| Reasoning | 84% | 92.5% | 89% | 88.5% |
| Multimodal Processing | 94.7% | 95% | 98% | 95.9% |
| Effort Reduction | 91% | 97% | 95% | 94.3% |
| **Overall** | **89.5%** | **95%** | **92.7%** | **92.4%** |

Task Accuracy: Specialists reported a proficiency level of 87%, with Generalists and Observers achieving even higher rates of 94% and 91.25%, respectively. On average, MSAM demonstrated a robust task accuracy rate of 90.75%, indicating its capability to deliver precise outcomes across diverse tasks.

Task Completeness: MSAM exhibited a high level of task completeness, with Specialists achieving a rate of 91%, followed closely by Generalists at 96.5%, and Observers at 90%. The collective average task completeness rate stood at 92.5%, highlighting MSAM's efficacy in executing tasks comprehensively.

Reasoning: Across all user cohorts, MSAM demonstrated notable reasoning abilities, as evaluated by humans, with respective scores of 84%, 92.5%, and 89% for Specialists, Generalists, and Observers. The average reasoning score of 88.5% underscores MSAM's capacity to facilitate logical decision-making processes within enterprise workflows.

Multimodal Processing: MSAM exhibited robust capabilities in processing (read/write) multimodal data formats, with high accuracy rates observed across all user cohorts. Specialists, Generalists, and Observers achieved impressive rates of 94.7%, 95%, and 98%, respectively, resulting in an average accuracy rate of 95.9%.

Effort Reduction: Users reported significant reductions in effort facilitated by MSAM, with Generalists recording the highest reduction rate of 97%, followed by Observers at 95%, and Specialists at 91%. On average, MSAM enabled a substantial effort reduction of 94.3%, indicative of its ability to streamline workflows and enhance operational efficiency.

Moreover, user feedback underscored the transformative impact of Empsing's deployment, with 75% of users attributing automation, effort reduction, and cost savings to MSAM's functionalities. While other 25% rated that the platform required additional effort to automate. Additionally, organizations leveraging Empsing observed an average reduction of 87% in project and departmental costs and efforts, attesting to the tangible benefits conferred by Empsing/MSAM's deployment.

Furthermore, proactive measures undertaken by the Empsing team resulted in the detection and resolution of 84 programmatic bugs and errors identified during user interactions, alongside the identification and mitigation of over 14 notable security vulnerabilities, affirming Empsing's commitment to delivering a secure and reliable enterprise AI solution.

**VII. BENCHMARK TESTING RESULTS**

Empsing's performance was rigorously evaluated against industry-leading models, including GPT-4 and Gemini, across various benchmark tests. The comparative analysis revealed Empsing's unequivocal task execution superiority in terms of intelligence and attention capacity, positioning it as the leader in intelligent enterprise problem-solving. Utilizing a robust evaluation framework, Empsing's performance metrics were juxtaposed against its counterparts, resulting in a definitive demonstration of its dominance in critical evaluation criteria.

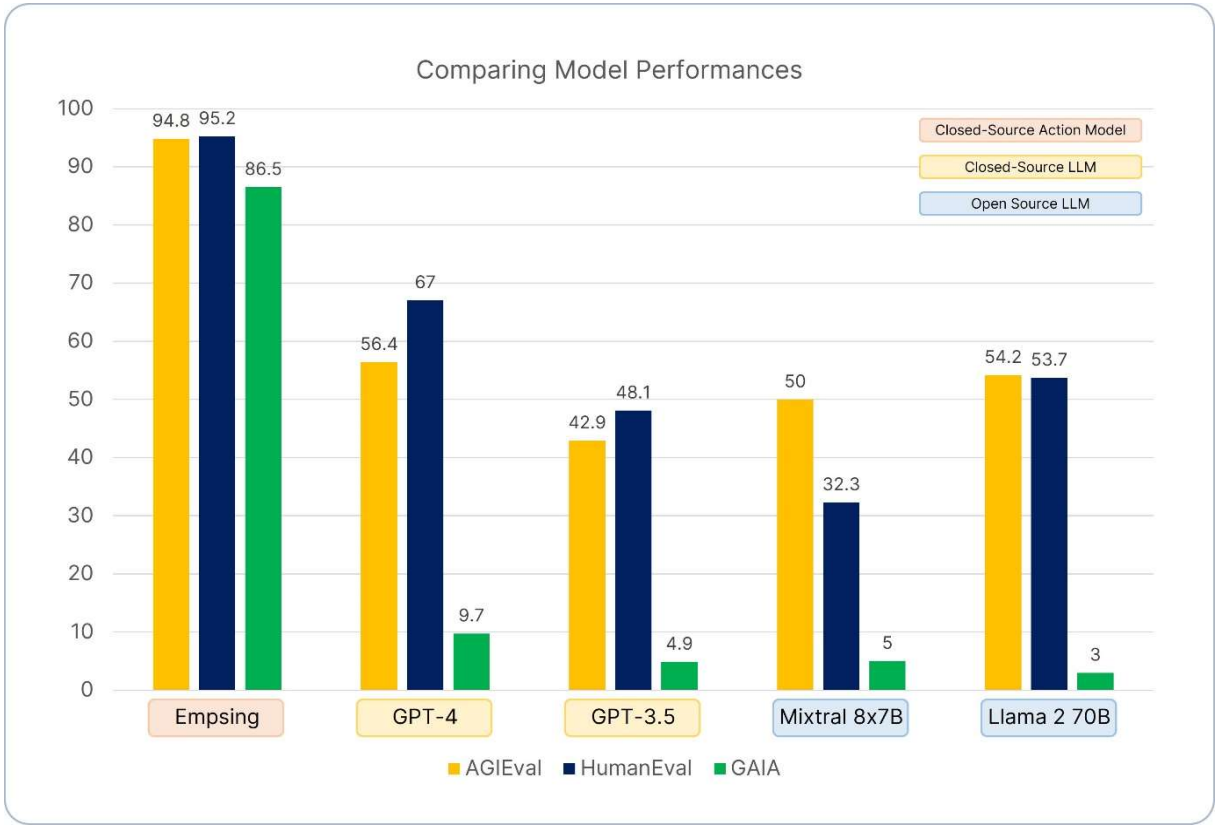

### AGIEval

Empsing garnered an impressive score of 94.8 in AGIEval, surpassing all other models. This signifies Empsing's exceptional proficiency in tasks associated with Artificial General Intelligence (AGI), underscored by its advanced capabilities in understanding and addressing complex challenges.

### GAIA

Empsing demonstrated superior performance with a weighted average score of 86.48% (and 78% unweighted). It scored 96% in Level 1, 94% in Level 2 and 44% in Level 3 of the GAIA benchmark. These results demonstrate MSAM outperforming major systems including Copilot (GPT-4 + Plugins), FRIDAY and GPT-4 Turbo. It also highlights MSAM's efficiency in handling tasks specific to General Intelligence, indicative of its capacity to provide accurate and effective solutions across diverse enterprise scenarios.

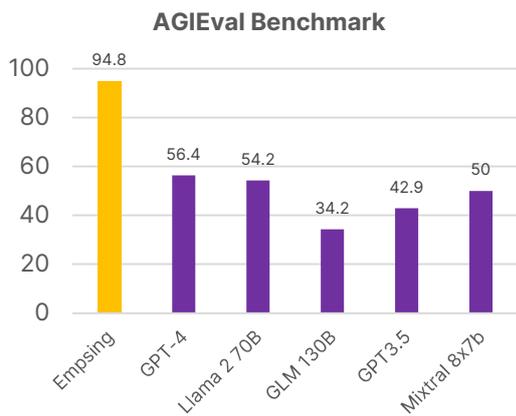

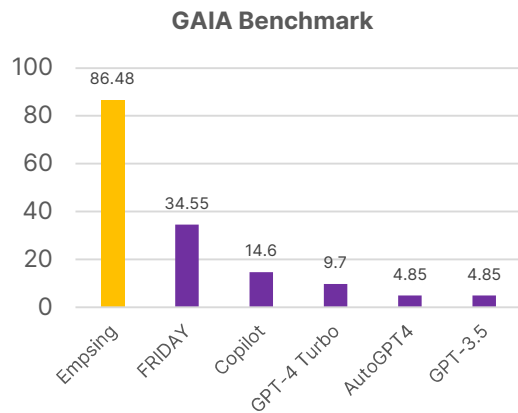

HumanEval

Empsing excelled with an outstanding score of 95.2 in HumanEval, surpassing all other models by a margin. This underscores Empsing's remarkable aptitude in tasks requiring human-like comprehension and reasoning abilities in coding, consolidating its position as a task-based model designed to operate as an employee.

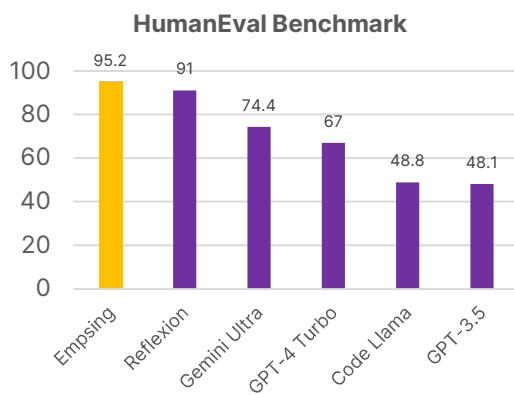

Through its exceptional performance across diverse benchmark tests, Empsing reports a new performance standard in AI-based Models, offering unparalleled capabilities in addressing complex challenges and advancing the frontiers of AI technology through its Digital Employees Platform.

## VIII. LIMITATIONS

Despite its considerable strengths and capabilities, the Multi-Step Action Model (MSAM) is not immune to certain limitations that warrant acknowledgment.

**Data Dependency:** MSAM's efficacy is contingent upon the availability and quality of data for organizational fine-tuning. In scenarios where relevant data is scarce or of insufficient quality, MSAM's performance may be generic, potentially impeding its effectiveness in addressing complex organizational tasks.

**Computational Resources:** The deployment and operation of MSAM necessitate substantial computational resources, including high-performance hardware and scalable infrastructure. Consequently, organizations with limited computational capabilities may encounter challenges in implementing MSAM on-premise at scale, thereby restricting its accessibility. However, they can still subscribe to Empsing's SaaS version or opt for a hybrid deployment.

**Complexity of Integration:** Integrating MSAM within existing enterprise ecosystems may pose challenges due to compatibility issues, legacy system constraints, and organizational resistance to change. The intricate nature of enterprise integration processes necessitates scrupulous planning, resource allocation, and stakeholder engagement to ensure seamless deployment and functionality.

**Domain Specificity:** While MSAM demonstrates versatility across diverse operational domains, its performance may vary based on the specificity and complexity of the task at hand. Certain specialized domains or niche industries such as Cancer Surgeries or use of Heavy Machinery may require tailored adaptations or additional training to optimize MSAM's performance, potentially limiting its applicability in certain contexts.

By addressing these challenges through proactive measures and strategic initiatives along with Empsing, organizations can leverage MSAM's strengths to enhance operational efficiency, decision-making processes, and overall business outcomes.

## IX. FUTURE OF MSAM

The trajectory of Empsing's Multi-Step Action Model (MSAM) is poised for continued innovation and evolution, heralding a future characterized by enhanced capabilities, expanded applications, and greater impact in driving enterprise AI adoption.

Future iterations of MSAM by Empsing will prioritize adaptability and scalability, enabling seamless integration within diverse enterprise ecosystems and accommodating evolving user requirements. Leveraging advancements in AI Infrastructure, Knowledge Protocols and Edge Computing, MSAM will exhibit greater flexibility and agility in addressing dynamic & specialized business needs across various domains.

Building upon its foundation of multi-step reasoning and task-based action, future iterations of MSAM will incorporate advanced cognitive capabilities, including enhanced comprehension, contextual reasoning, and personalized decision-making. By leveraging state-of-the-art techniques, MSAM will exhibit unprecedented levels of cognitive sophistication, enabling it to navigate complex scenarios and adapt to evolving user preferences with greater precision and efficacy.

## X. CONCLUSION

The Multi-Step Action Model (MSAM) represents a pivotal advancement in fostering enterprise-readiness for AI adoption through Empsing, offering a versatile platform rooted in accuracy, adaptability, scalability, and ethical AI practices. Through meticulous design and rigorous evaluation, MSAM has demonstrated its efficacy in addressing complex business challenges, exhibiting superior performance in intelligence quotient, attentiveness, and overall functionality.

As organizations navigate the complexities of the AI landscape, Empsing and its core – MSAM – can emerge as a strategic enabler of intelligent automation and responsible AI deployment. By prioritizing continuous research, collaboration, and ethical stewardship, enterprises can harness the transformative potential of MSAM to drive innovation, productivity, and competitiveness in the modern business environment.